\documentclass[graybox]{SNmult}

\usepackage{type1cm}        
\usepackage{cite}
\usepackage{makeidx}         
\usepackage{graphicx}        
\usepackage{multicol}        
\usepackage{amsfonts, mathtools}
\usepackage[bottom]{footmisc}
\usepackage[breaklinks=true,colorlinks,allcolors=blue]{hyperref}
\usepackage{comment}

\usepackage{newtxtext}       %
\usepackage[varvw]{newtxmath}       

\usepackage{braket}

\newcommand{\trace}[1]{\text{tr}[#1]}

\newcommand{\proj}[1]{\ket{#1}\!\bra{#1}}
\newcommand{\eqlabel}[1]{\eqref{#1}}
\newcommand{\figlabel}[1]{Fig.~\ref{#1}}
\newcommand{\seclabel}[1]{Sect.~\ref{#1}}

\newcommand{\hc}{\text{h.c.}}

\makeindex

\begin{document}
\title*{Bridging continuous control and Floquet driving for charging many-body spin chains}
\author{Sebastián V. Romero, Xi Chen and Yue Ban
}

\institute{Sebastián V. Romero \at Quantum Advanced Research Center (QuARC), CSIC, 28049 Madrid, Spain \at Instituto de Ciencia de Materiales de Madrid (ICMM), CSIC, 28049 Madrid, Spain \at Departamento de Física Teórica de la Materia Condensada, Universidad Autónoma de Madrid, 28049 Madrid, Spain \at \email{sebastian.v.romero@csic.es}
\and Xi Chen \at Quantum Advanced Research Center (QuARC), CSIC, 28049 Madrid, Spain \at Instituto de Ciencia de Materiales de Madrid (ICMM), CSIC, 28049 Madrid, Spain \at \email{xi.chen@csic.es}
\and Yue Ban \at Quantum Advanced Research Center (QuARC), CSIC, 28049 Madrid, Spain \at Instituto de Ciencia de Materiales de Madrid (ICMM), CSIC, 28049 Madrid, Spain \at \email{yue.ban@csic.es}}

\maketitle

\abstract{
Recent advances in quantum information and quantum thermodynamics have reshaped the understanding of energy storage at the microscopic scale, paving the way toward protocols for storing and transferring energy in quantum devices. These systems, known as quantum batteries, offer a conceptual alternative to conventional macroscopic chemical batteries by exploiting quantum coherence, correlations, and many-body dynamics. By navigating the landscape of established spin-based quantum batteries, we review existing charging and work-extraction protocols, as well as the impact of external factors on their performance. Motivated by Floquet-engineered proposals, we further establish a connection between continuously and periodically driven spin chains as platforms for quantum energy storage. Finally, we survey experimental realizations and proposals, highlighting their implementability and scalability in current and near-term quantum technologies.
}

\section{Introduction}\label{sec:intro}

Quantum batteries (QBs) have emerged as a vibrant interface between quantum thermodynamics, many-body physics, and quantum information theory~\cite{alicki2013entanglement,hovhannisyan2013entanglement,binder2015quantacell}. A QB is defined as a quantum system that can store and supply energy in the form of work. The motivation for studying QBs is twofold. On the one hand, they offer a clean setting in which to revisit the notion of work in quantum mechanics and to identify how much of it is actually extractable. On the other hand, they are relevant to future quantum technologies, where on-chip controlled power supply and fast and efficient storage may become essential. In this setting, the central issue is not merely how much energy can be deposited into a system, but how much of it can later be extracted. While most work to date has been theoretical, there have been significant experimental efforts to realize QBs in various platforms, including superconductors~\cite{hu2022optimal,hu2026quantumcharging}, quantum dots~\cite{wenniger2023experimental}, organic microcavities~\cite{quach2022superabsorption} and nuclear spins~\cite{joshi2022experimental}.

In contrast to their classical counterparts, QBs leverage quantum coherence and many-body operations to generate non-classical correlations, which can yield a genuine quantum advantage in the charging process. In other words, these might lead to the so-called superextensive scaling of the charging power, meaning that this quantity can grow faster than linearly with the number of battery cells. This goes beyond the best classically attainable behavior, where the charging power scales at most extensively with system size. Achieving superextensive scaling has been a major motivation for the study of QBs, as it suggests that quantum effects can be harnessed to outperform classical energy storage devices. However, the precise conditions under which such an advantage can be realized remain active areas of research, where early work has elucidated under which conditions such advantage can be attained~\cite{campaioli2017enhancing,farre2020bounds,gyhm2022quantum}.

Recent studies have developed theoretical frameworks where a scaling advantage might be reached, including a wide variety of setups, as for example the quantum analogue of the Dicke model~\cite{ferraro2018highpower} and non-linearities in harmonic oscillators~\cite{andolina2025genuine}. Among them, Le \emph{et al.} proposed in 2018 chains of spins as many-body QBs~\cite{le2018spinchain}, whose deceptively simple definition can lead to superextensive scaling thanks to collective operators via long-range interactions. This foundational work has inspired a plethora of subsequent studies on spin-chain batteries~\cite{rossini2019manybody, ghosh2020enhancement, arjmandi2022enhancing, ghosh2022dimensional, ali2024ergotropy, bhattacharya2026heisenberg, huangfu2021highcapacity, gao2022scaling, grazi2024controlling, catalano2024frustrating, dou2022charging, schmid2026superextensive, ho2026boosting, grazi2025charging, pavone2026cluster, zhao2025nonmarkovian, dou2022cavity, sun2025cavity, zhao2021quantum, chang2021optimal, yao2022optimal, fajar2025noise, hu2026quantumcharging, liu2021entanglement, peng2021lower, qi2021magnon, barra2022quantum, guo2024analytically}. While most of these works have continuously-driven local fields present, recent proposals have considered infinitesimally narrow pulses as transverse fields instead, leading to kicked-Ising-based models~\cite{romero2025kicked,mondal2022periodically,pg2025dichotomy,puri2025floquet,romero2026impact}. These models encapsulate diverse enriching dynamical properties, such as quantum chaos and many-body localization, as well as a new class of non-equilibrium phases, as the discrete time crystalline phases~\cite{zaletel2023quantum}.

More recent proposals have considered the Sachdev-Ye-Kitaev (SYK) model, an all-to-all connected model of quantum holography displaying an approximate conformal symmetry and fast scrambling at low temperatures~\cite{sachdev1993gapless,kitaev2015talk,maldacena2016remarks}, as a QB~\cite{rosa2020ultrastable,rossini2020quantum,kim2022operator,romero2025scrambling,divi2025syk}. The SYK model extends spin-chain setups by replacing spin-spin interactions with all-to-all random interactions. Despite being a maximally chaotic model, it has been shown that the charging advantage of SYK batteries is not directly related to their chaotic properties, but rather to the structure of their interactions and the resulting collective dynamics~\cite{romero2025scrambling}. This model allows one to ask whether the collective charging advantage present in spin chains persists when integrability and spatial order are absent. 

This Chapter develops a focused perspective on spin-chain and kicked-Ising QBs, which provide a natural setting to clarify the connection between many-body interactions, continuous control, and Floquet driving. Within this scope, we discuss how many-body interactions, collective quantum correlations, and non-equilibrium driving protocols shape the charging performance of a QB. The central thread is the bridge between continuous and Floquet charging protocols in many-body spin chains, with particular emphasis on the role of interaction range, disorder, and periodically kicked dynamics. We also include SYK QBs as a complementary all-to-all random-interaction setting, which allows us to assess whether collective charging advantages persist beyond spatially ordered spin-chain models.
\begin{figure}[!tb]
    \centering
    \includegraphics[width=\linewidth]{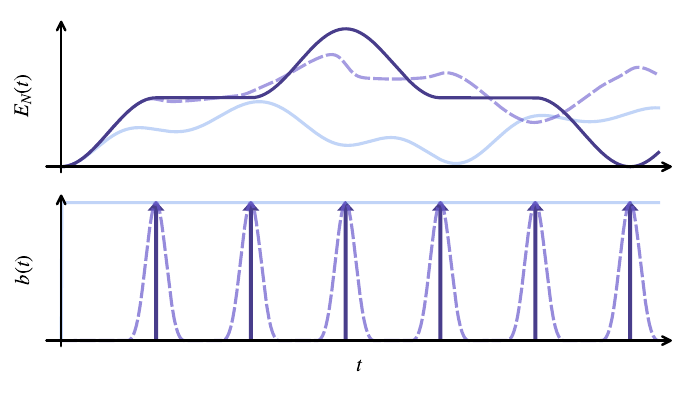}\vspace{-4mm}%
    \caption{Schematic of an Ising-chain QB bridging continuously driven and periodically kicked transverse-field charging protocols. From lighter to darker blue, the injected energy in a QB (top) under charging protocols with decreasing pulse width (bottom). While a continuous drive (represented by a constant line) yields a poor performance, a periodically-kicked pulse (depicted with arrows) reaches maximal charging without fluctuations. Between these two limiting cases, a finite-width periodic pulse coined as \emph{quasikick}~\cite{romero2025kicked} (dashed line), displays a performance lying between both limit regimes.}
    \label{fig:schematic}
\end{figure}%

The rest of the Chapter is organized as follows. While in~\seclabel{sec:preliminaries} we introduce the basic QB framework and the figures of merit used to assess the charging performance, in~\seclabel{sec:spin} we discuss the main results on spin-chain batteries, including the role of disorder and long-range interactions and anisotropy. In~\seclabel{sec:floquet} we introduce the kicked-Ising model as a Floquet-engineered QB, showing how it can be solved exactly and how it can be used to establish a link between continuously- and periodically-driven spin chains for energy storage (see~\figlabel{fig:schematic} for a schematic representation). Additionally, in~\seclabel{sec:syk} we briefly review SYK batteries, including their scaling advantage and the role of chaos in charging performance. Finally, in~\seclabel{sec:discussion}, we summarize the main conclusions and outline promising directions for future research.

\section{Quantum battery framework and figures of merit}\label{sec:preliminaries}

In this section, we introduce the basic framework of QB and relevant notations used throughout this chapter to assess the charging performance of QBs. Hereafter, we make use of natural units and set $\hbar\equiv k_\text{B}\equiv 1$ accordingly.

Following the foundational work of Alicki and Fannes~\cite{alicki2013entanglement}, a QB is a $d$-dimensional quantum system $H_\text{b}$ which is capable of storing energy by driving the system toward an excited state $\rho$ such that the mean energy injected satisfies that $\trace{\rho H_\text{b}} > \varepsilon_1$, with spectral decomposition $H_\text{b} = \sum_{j=1}^d \varepsilon_j \proj{\varepsilon_j}$, with $\varepsilon_j\le\varepsilon_{j+1}$ $\forall j$ and $\varepsilon_d-\varepsilon_1>0$. This energy injection can be realized either through a unitary process or a time evolution involving incoherent contributions, thus {an open system}. 

For a closed QB, the total Hamiltonian is usually written as $H(t) = H_\text{b} + V(t)$, with $V(t)$ an external driving Hamiltonian acting on the battery Hamiltonian $H_\text{b}$ within a charging time window $t\in[0,\tau]$. The state of the battery then evolves according to the von Neumman equation 
\begin{equation}\label{eq:closed}
    \frac{\text{d}\rho(t)}{\text{d}t} = -i [H(t), \rho(t)],
\end{equation}
where $\rho(t) = U(t)\rho(0)U^\dagger(t)$ is the time-evolved system state with $\rho(0)$ the initial state, usually taken to be the ground state of $H_\text{b}$ (battery discharged), and $U(t) = \mathfrak{T}\exp\big[-i\int^t_0 \text{d}t' H(t')\big]$ is the time-evolved unitary operator with $\mathfrak{T}\exp[\cdot]$ being the time-ordering. When the system is weakly coupled to an environment whose correlation time is much shorter than the characteristic system-evolution time, the open part of the dynamics can be described, within the Born-Markov approximation, by Lindblad form through jump operators $L_i$, that is, extending~\eqlabel{eq:closed} into a Lindblad master equation,
\begin{equation}\label{eq:lindblad}
  \frac{\text{d}\rho(t)}{\text{d}t} = -i[H(t), \rho(t)] + \sum_i \left[ L_i\rho(t)L^\dagger_i - \frac{1}{2}\left\{L^\dagger_iL_i,\rho(t)\right\} \right].
\end{equation}

In the following, we focus primarily on direct charging protocols, where the system is generally decomposed as $H(t) = H_\text{b} + \lambda(t)H_\text{c}(t)$, with $\lambda(t)$ a switching function that toggles between the battery Hamiltonian $H_\text{b}$ and charging Hamiltonian $H_\text{c}(t)$, respectively, which is usually taken as a step function for simplicity, with an amplitude equal to one along $t\in[0,\tau]$ and zero otherwise. To ensure energy injection, both Hamiltonians have to satisfy that $[H_\text{b}, H_\text{c}(t)]\neq 0$. This protocol assumes that the charger is an infinite energy reservoir, which differs from the so-called charger-mediated protocol, where the energy is transferred from one charged battery to another one discharged through an interaction Hamiltonian between both systems~\cite{campaioli2024colloquium}.

The energy injected after charging completion is $E(\tau)=\trace{\rho(\tau)H_\text{b}} - \trace{\rho(0)H_\text{b}}$, where the optimal charging time $\tau^*$ is usually defined as the time where the mean charging power $P(\tau)=E(\tau)/\tau$ peaks, namely $\tau^*=\arg \max_\tau P(\tau)$. This maximum is a key metric for assessing whether a protocol features or not a charging advantage: parameterizing the maximum charging power scaling with system size as $P(\tau^*)\sim N^{1+\ell}$, a charging protocol is said to feature a quantum charging advantage if $0<\ell\le 1$~\cite{gyhm2022quantum}, known as superextensive scaling, which can not be attained by any classical counterpart, which can achieve at most an extensive scaling $(\ell=0)$. The pursue of charging protocols exhibiting a superextensive scaling has been one of the main driving forces in the QB community, where this advantage is said to be genuine when it is originated from quantum correlations, but its definition is still under debate~\cite{campaioli2024colloquium}.

From a practical perspective, the injected energy alone does not fully quantify the usefulness of a charged QB.  A more direct measure of extractable work is  the ergotropy~\cite{allahverdyan2004maximal}, defined as the maximum amount of work that can be extracted from a given state $\rho$ through unitary operations. Considering the spectral decomposition of the state as $\rho = \sum_{j} r_j \proj{r_j}$ ($r_j \ge r_{j+1}$), the ergotropy is defined as 
\begin{equation}\label{eq:ergotropy}
    \mathcal{W}(\rho) = \trace{\rho H_\text{b}} - \trace{\rho_{\text{p}} H_\text{b} },
\end{equation} 
with $\rho_{\text{p}} = \sum_{j} r_j \proj{\varepsilon_j}$ defined as the passive state associated with $\rho$~\cite{pusz1978passive}, characterized by populations that do not increase with energy, thus yield zero ergotropy. Hence, ergotropy is always lower or equal than the energy injected.%

Finally, practical QBs must satisfy stability and retention requirements. The stored energy should not exhibit strong temporal fluctuations that require decoupling the charger from the battery in unrealistic time windows to prevent immediate energy backflow and mitigating self-discharging, so that the stored energy does not leak into the environment over time once the charger is decoupled~\cite{santos2019stable}.

\section{Spin-chain batteries}
\label{sec:spin}

Spin chains, among the most deceptively yet versatile many-body quantum systems encapsulating a wide range of phenomena, have been widely explored as QBs. One reason for their usefulness
is that several paradigmatic one-dimensional models, including the transverse-field Ising and XY chains, admit an exact mapping to free fermions. In these free-fermionic cases, the Jordan--Wigner transformation maps the spin Hamiltonian onto a quadratic fermionic form, enabling a controlled analytical
description of the charging dynamics. The resulting Hamiltonian can be written
as~\cite{lieb1961two}
\begin{equation}\label{eq:quadratic}
    H_\text{c} = \sum_{i,j=1}^N \left[ c^\dagger_i A_{ij} c_j + \frac{1}{2}\left( c^\dagger_i B_{ij} c^\dagger_j + \hc \right) \right],
\end{equation}
with $c^\dagger_i$ ($c_i$) a spinless creation (annihilation) fermion at the $i$th site, following the anticommutation rules $\{c_i,c^\dagger_j\}=\delta_{ij}$ and $\{c_i,c_j\}=0$, and $A$ and $B$ encoding the hopping and pairing terms, respectively. The spin-$1/2$ representation of~\eqlabel{eq:quadratic} can be obtained via the Jordan-Wigner transformation $c^\dagger_j = -\sigma^-_j\left( \prod_{k=1}^{j-1} \sigma^z_k \right)$ with $\sigma^\pm_j = \left(\sigma^x_j \pm i\sigma^y_j \right)/2$ the ladder operators at site $j$.

In 2018, Le \textit{et al.}~\cite{le2018spinchain} proposed the first spin-chain QB model, with the battery and charging Hamiltonians ($H_\text{b}$ and $H_\text{c}$, respectively), given by 
\begin{equation}\label{eq:le_et_al}
    H_\text{b} = B \sum_{i=1}^N \sigma_i^z - \sum_{i<j} g_{ij} \left[ \sigma^z_i \sigma^z_j + \alpha\left( \sigma^x_i \sigma^x_j + \sigma^y_i \sigma^y_j \right) \right], \quad H_\text{c} = \omega \sum_{i=1}^N \sigma_i^x.
\end{equation}
Here, $B$ and $\omega$ set the strength of the external Zeeman and transverse fields, respectively.
$g_{ij} = g/|i-j|^p$ is the power-law-decaying interaction strength, while $\alpha$ controls the anisotropy (with $|\alpha|\le 1$). The choices $\alpha=0$ and
$\alpha=1$ recover the Ising and isotropic Heisenberg (XXX) limits,
respectively, whereas intermediate values describe anisotropic XXZ-type
interactions. Particularly, in this proposal the interactions modify the spectrum and eigenstates in which energy is stored, even though the external charging drive itself is local. The resulting collective response depends strongly on the interaction range, whose impact on the charging power scaling with respect to the system size ranges from an extensive scaling for the nearest-neighbor limit ($p\to\infty$) and a super-extensive scaling $P_N\sim N^2$ for the infinite-range case ($p=0$). In fact, the intermediate $p=1$ scenario is studied, returning a more moderate superextensive scaling $P_N\sim N\log N$ with respect to the $N^2$ limit where uniform couplings are present for all the possible $(i,j)$ pairwise interactions among the spins.

From~\eqlabel{eq:le_et_al} we can observe that the battery Hamiltonian $H_\text{b}$ is a many-body Hamiltonian when $g_{ij}\neq0$, where the energy is stored in the interactions among spins, while the charging Hamiltonian $H_\text{c}$ is a sum of local transverse field terms, which differs from standard approaches where the energy is stored in the local terms of the Hamiltonian, and the charging is performed through many-body interactions instead. When interactions are not present $(g_{ij}=0)$, an oscillatory behaviour of the energy injected is expected, which is highly undesirable for practical applications~\cite{santos2019stable}.

Interestingly, the role of the anisotropy $\alpha$ is also discussed, where the mere presence of many-body interactions does not necessarily guarantee an enhancement in the charging power. In the isotropic
case $\alpha=1$, corresponding to the XXX model, the interaction part of the
battery Hamiltonian is rotationally invariant and commutes with the collective
charging Hamiltonian $H_\text{c}$. Consequently, the interaction energy remains unchanged under the charging
drive, and only the local Zeeman contribution is dynamically charged. The
resulting maximum power therefore retains the classically attainable extensive
scaling, $P_N(t)\sim N$. By contrast, as the anisotropy is varied from
$\alpha=1$ (XXX chain) toward $\alpha=-1$ (XXZ chain), the rotational symmetry is broken and the
interaction becomes dynamically active during the charging process. The
charging power is consequently enhanced, reaching a substantially larger value
in the anisotropic $\alpha=-1$ limit.
These results demonstrate that exploiting
many-body interactions for collective charging requires not only their presence
but also the breaking of the rotational symmetry that otherwise renders them
inactive under the charging drive.

Therefore, the presence of many-body interactions is not a sufficient condition to achieve a quantum charging advantage, but rather the structure of the interactions and the resulting collective dynamics play a crucial role in determining the charging performance. In this regard, the interplay between the interaction range and the anisotropy parameter $\alpha$ is key to understanding how to optimize the charging process in spin-chain batteries.

This foundational work by Le \emph{et al.} has inspired a plethora of subsequent studies on spin-chain batteries, including solid-state QBs~\cite{hu2026quantumcharging}, disordered systems~\cite{rossini2019manybody,ghosh2020enhancement,arjmandi2022enhancing}, Heisenberg chains and charging protocols based on Dzyaloshinskii-Moriya interaction~\cite{ghosh2022dimensional,ali2024ergotropy,bhattacharya2026heisenberg}, XY chains~\cite{huangfu2021highcapacity,gao2022scaling,grazi2024controlling}, frustrated QBs~\cite{catalano2024frustrating}, Lipkin-Meshkov-Glick models~\cite{dou2022charging,schmid2026superextensive,ho2026boosting}, cluster-Ising models~\cite{grazi2025charging,pavone2026cluster}, thermal charging~\cite{zhao2025nonmarkovian}, cavity-assisted charging~\cite{dou2022cavity,sun2025cavity}, charging protocols assisted through an open evolution~\cite{zhao2021quantum,chang2021optimal,yao2022optimal,fajar2025noise}, central-spin QB models~\cite{liu2021entanglement,peng2021lower}, magnon-mediated charging processes~\cite{qi2021magnon}, and Ising chain models~\cite{barra2022quantum,guo2024analytically}, among many others.

\section{Floquet batteries}
\label{sec:floquet}

We now turn from continuously driven protocols to periodically driven charging. In a kicked protocol, part of the Hamiltonian is applied through short pulses that are ideally represented by Dirac delta functions (the so-called \emph{kicks}). The dynamics are therefore naturally described stroboscopically through a Floquet operator that propagates the system over one driving period. Despite this discrete-time description, the evolution remains fully coherent in the absence of dissipation. Periodically driven many-body systems exhibit a broad range of dynamical phenomena, including time-crystalline phases~\cite{zaletel2023quantum}, quantum chaos~\cite{rozenbaum2017lyapunov}, many-body localization~\cite{waltner2021localization}, and particle-time duality~\cite{Akila_2016}. These features motivate Floquet QBs, in which charging is implemented through periodic pulses rather than a continuously applied drive.

Among the different proposals present in literature~\cite{romero2025kicked,romero2026impact,mondal2022periodically,pg2025dichotomy,puri2025floquet}, we focus on the kicked-Ising QB, whose charging Hamiltonian is described by
a periodically kicked transverse-field Ising model, also known as the kicked-Ising chain (KIC). This features
maximal charging at selected Floquet cycles, together with robustness against disorder~\cite{romero2025kicked}. The complete charging protocol is generated
by $H(t)=H_\text{b} + \lambda(t)[H_\text{c}(t) - H_\text{b}]$, where $\lambda(t)=1$ for $t\in[0,\tau]$ and vanishes otherwise. During the
charging interval, the charger Hamiltonian is $H_\text{c}(t) = H_\text{I} + H_\text{K}\sum_{m\in\mathbb{Z}} \delta(t-mT)$, with Ising and kicked transverse field contributions,
\begin{equation}\label{eq:kic}
    H_\text{I} = \sum_{\braket{ij}} J_{ij} \sigma_i^x \sigma_j^x + \sum_{i=1}^N h_i \sigma_i^x, \qquad H_\text{K} = \sum_{i=1}^N b_i \sigma^z_i.
\end{equation}
Here $J_{ij}$ is the interaction strength between spins $i$ and $j$, and $h_i$ and $b_i$ are the strengths of the longitudinal and transverse fields. In the following, we set a unit driving period $T=1$. For uniform parameters $J_{ij}=J$, $b_i=b$, and $h_i=0$, the self-dual
point is obtained for $J=\pi/4$ and $b=-\pi/4$. At this point, the Floquet
dynamics displays maximal entanglement growth and operator spreading~\cite{Akila_2016}, which might be desirable properties for a QB. The battery Hamiltonian is $H_\text{b}=(g/2)\sum_{i=1}^N \sigma^z_i$ and and the battery is initialized in its ground state (GS) and we set $g=1$ hereinafter.

During one driving period, the evolution is generated by the Floquet
operator
\begin{equation}\label{eq:floquet}
    U(1) = \mathfrak{T}\exp\left[ -i\int^1_0 \text{d}t H(t) \right] = e^{-i H_\text{K}} e^{-i H_\text{I}}.
\end{equation}
Observe that the Floquet operator can be interpreted as a stroboscopic evolution of the system, where the state of the system is updated at discrete time intervals corresponding to the period of the driving, obtaining that $U(m) = U^m(1)$. As a side note, periodically-kicked models are said to follow gate-like dynamics, where the exact evolution can be decomposed into a sequence of unitary gates, as shown in~\eqlabel{eq:floquet}. This gate-like structure makes kicked models natural candidates for benchmarking digital quantum hardware. In simulations of continuously driven Ising dynamics, the target evolution is approximated by a Trotterized circuit, which introduces discretization errors controlled by the resolution of the time steps~\cite{kim2023evidence,miessen2024benchmarking,visuri2026digitized}. By contrast, kicked-Ising chains and discrete time crystals are intrinsically stroboscopic. Since the Floquet evolution is written as a sequence of elementary unitary operations, the benchmark directly probes gate implementation, calibration, and noise errors rather
than Trotterization errors~\cite{romero2025kicked,fischer2026dynamical,shinjo2026unveiling}. 

The exact charging dynamics can be obtained through two complementary
analytical methods, summarized in~\figlabel{fig:kic_diag}. The first exploits the Clifford quantum cellular automata structure of~\eqlabel{eq:floquet} at the self-dual point, where the conjugation of a Clifford operator $U_\mathcal{C}$ over a Pauli string $O$ returns another Pauli string. For the particular case we are interested, where the Floquet operator is split into $U_\mathcal{C}=e^{-i\theta P}$ operators with $P$ a Pauli string, we have 
\begin{equation}\label{eq:clifford}
    e^{i\theta P} O e^{-i\theta P} = \begin{cases}
        O & \text{if }[O,P] = 0 \\
        \cos(2\theta) O - \frac{i}{2}\sin(2\theta) [O,P] & \text{otherwise}
    \end{cases}.
\end{equation}
Given that the battery Hamiltonian is a  sum of local Pauli strings, the charging dynamics can be captured by evolving in the Heisenberg picture each quantum cell of $H_0$, namely $\sigma^\alpha_i(m) = (U_KU_I)^{\dagger m} \sigma^\alpha_i (U_KU_I)^m$ with ($\alpha\in\{x,y,z\}$) and then compute via the induction $E_N(m) = \braket{\psi(m) | H_0 | \psi(m)} = (g/2)\braket{\psi(0) | \sigma^\alpha_i(m) | \psi(0)}$. 
While this first method holds regardless of the boundary conditions considered, it is limited to the self-dual point due to the need of a Clifford structure along the evolution. Nonetheless, it is still possible to analytically characterize for arbitrary $(J,b)$ pairs in~\eqlabel{eq:kic} when periodic boundary conditions (PBC) are considered. 

For the second method developed, when $h_i=0$ in~\eqlabel{eq:kic}, the model becomes integrable, which maps to the quadratic form in~\eqlabel{eq:quadratic} after Jordan-Wigner transformation. Therefore, after exactly diagonalizing the KIC through the standard combination of Jordan-Wigner, Fourier and Bogoliubov-de Gennes transformations, one lands into $2\times 2$ Floquet operators per momentum $k$, whose integer powers can be computed neatly via Chebyshev polynomials. More concretely, the system Hamiltonian reads after applying these transformations as
\begin{equation}\label{eq:kic_momentum}
\begin{split}
    H(t) = \sum_k \left[ \epsilon_k(t) c^\dagger_kc_k -\frac{\Delta_k}{2} (c^\dagger_kc^\dagger_{-k} + c_{-k}c_k) \right] = \sum_k \Psi^\dagger_k \mathcal{H}_k(t)\Psi_k,
\end{split}
\end{equation}%
where $\epsilon_k(t) = 2[J \cos k - b(t)]$, $b(t)=b\sum_{m\in\mathbb{Z}}\delta(t-m)$, $\Delta_k = 2J \sin k$ and constant contributions have been ignored. 
Here, $\mathcal{H}_k(t) = \epsilon_k(t) \tau^z - \Delta_k \tau^x$ is the Bogoliubov-de Gennes Hamiltonian in the Nambu basis $\Psi_k = [c_k, c_{-k}^\dagger]^{\text{T}}$, with eigenvalues $\lambda^\pm_k(t) = \pm E_k(t) = \pm \sqrt{\epsilon_k^2(t) + \Delta_k^2}$ and Pauli matrices $\tau^{x,y,z}$. Therefore, the system is diagonalized and split into $(k,-k)$ sectors. Their corresponding time evolution can be computed by solving the Schrödinger-like equation $i\text{d}U_k(t)/\text{d}t = \mathcal{H}_k(t) U_k(t)$, obtaining as Floquet operator after one kick
\begin{equation}\label{eq:floquet_momentum}
\begin{split}
    U_k(1) &= \mathfrak{T}\exp\left[ -i\int^1_0 \text{d}t \mathcal{H}_k(t) \right] \\
    &= \exp\left( 2ib\tau^z \right) \exp \left[ -2iJ(\cos k\tau^z - \sin k\tau^x) \right] = \begin{bmatrix}\alpha_k & -\beta^*_k\\\beta_k & \alpha^*_k\end{bmatrix},
\end{split}
\end{equation}
satisfying again that $U_k(m)=U^m_k(1)$, powers that can be computed via degree-$m$ Chebyshev polynomials of the second kind in $x$, $\mathcal{U}_m(x=\cos\theta)=\sin(m+1)\theta/\sin\theta$, as $U^m_k(1) = U_k(1)\mathcal{U}_{m-1}(\xi_k) - \mathcal{U}_{m-2}(\xi_k)$ with $\xi_k=\trace{U_k(1)}/2 = \text{Re}[\alpha_k]$.
\begin{figure}[!tb]
    \centering
    \includegraphics[width=.5\linewidth]{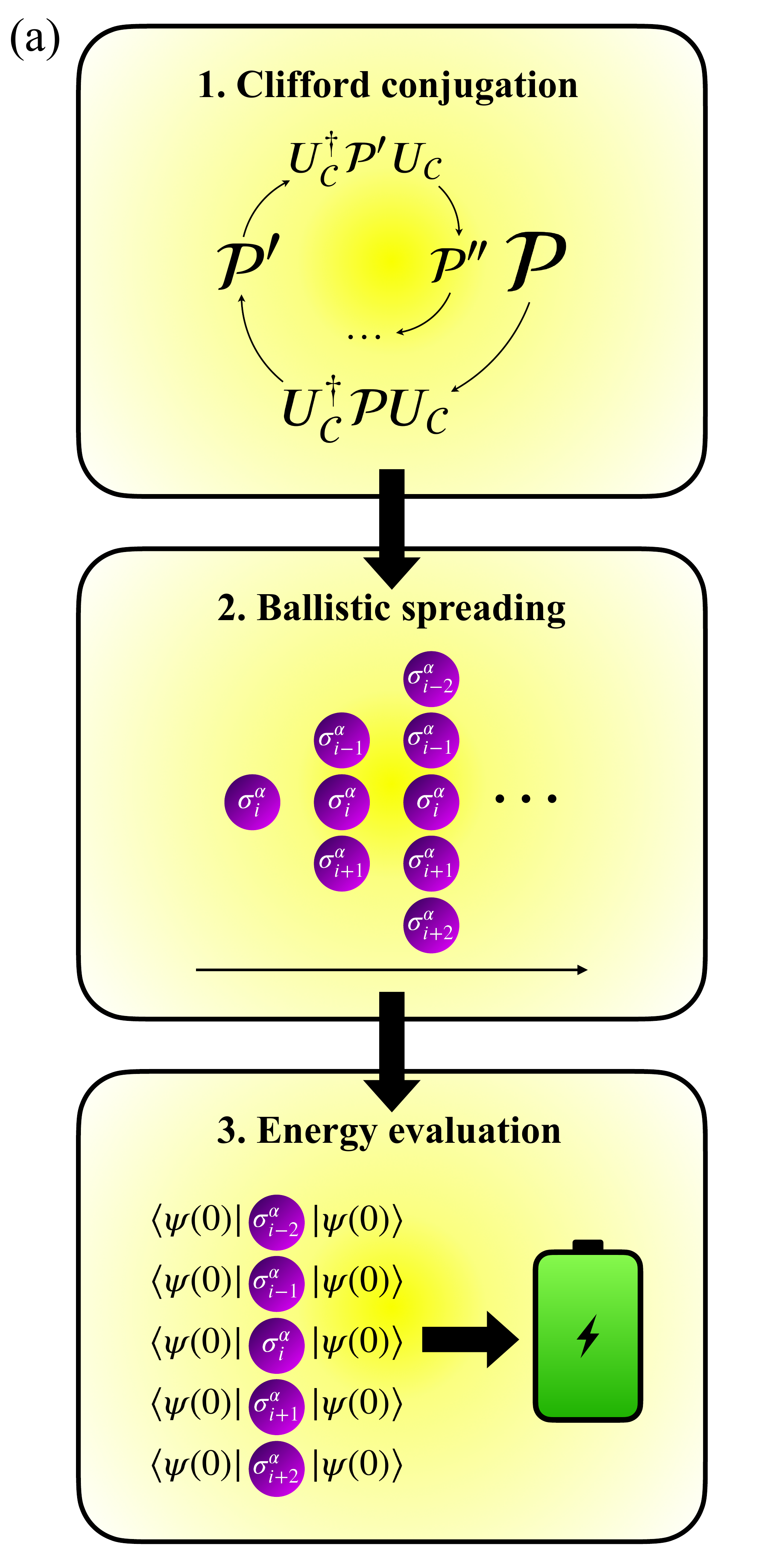}%
    \includegraphics[width=.5\linewidth]{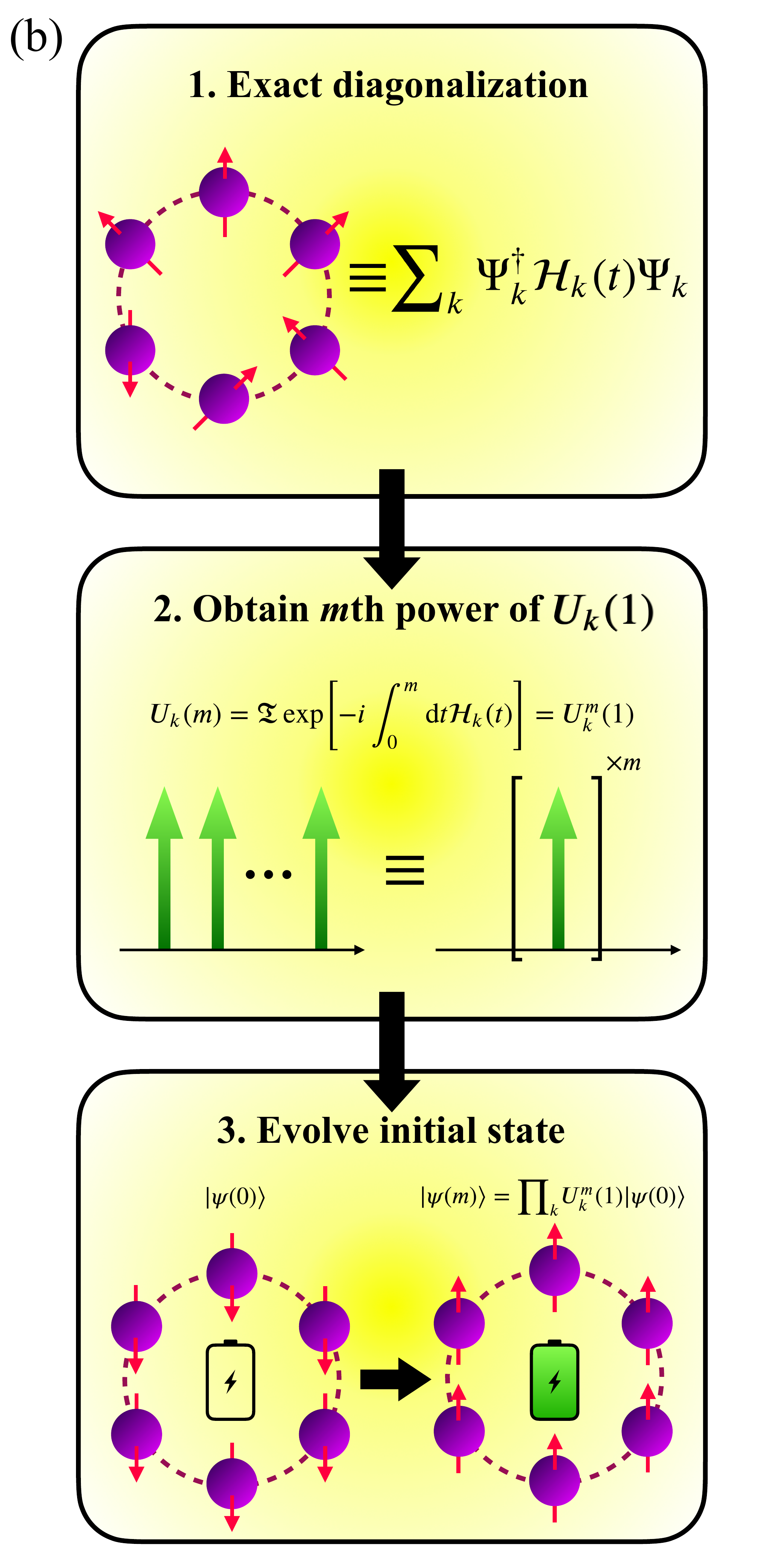}\vspace{-2mm}
    \caption{Analytical methods derived for the KIC QB. (a) For the Clifford quantum cellular automata: 1) at the self-dual point, the system exhibits a Clifford structure, where Clifford conjugation maps Pauli strings to Pauli strings (see~\eqlabel{eq:clifford}); 2) apply repeatedly the Clifford conjugation $U(1)$ in~\eqlabel{eq:floquet} to each cell, whose Heisenberg picture feature a ballistic operator growth, and 3) since the initial state is a product state, the energy injected is a sum of measured time-evolved cells per site. (b) To solve the kicked-Ising model in momentum space: 1) diagonalize the system by applying consecutively the Jordan-Wigner transformation and Fourier transform, 2) integrate the resulting Bogoliubov-de Gennes equations (see~\eqlabel{eq:floquet_momentum}) and compute the evolution after $m$ kicks via Chebyshev polynomials noting that $U_k(m)=U^m_k(1)$, and 3) the state is reconstructed as $\ket{\psi(m)} = \prod_k U_k(m)\ket{\psi(0)}$, yielding $E_N(m) = \braket{\psi(m) | H_0 | \psi(m)}$.}\label{fig:kic_diag}
\end{figure}%

In this setup, the initial state decomposes as $\ket{\psi(0)} = \prod_k \ket{\psi_k(0)} = \prod_k \ket{\downarrow}$ per momentum, the time-evolution can be computed as $\ket{\psi(m)} = \prod_k U^m_k(1) \ket{\psi_k(0)}$. Therefore, the final energy can be computed as
\begin{equation}
\begin{split}
    E_N(m) &= \frac{g}{2} \sum_k \big( \braket{\psi_k(m)|\sigma^z|\psi_k(m)} - \braket{\psi_k(0)|\sigma^z|\psi_k(0)} \big) \\
    &= g\sum_k |\alpha_k|^2\mathcal{U}^2_{m-1}(\xi_k) = g\sin^2(2J)\sum_k \sin^2 k\frac{\sin^2 m\theta_k}{\sin^2 \theta_k},
\end{split}
\end{equation}
which simplifies to $E_N(m) = g\sum_k \sin^2 mk$ at the self-dual point. 

Combining both analytical methods and $q\in\mathbb{Z}$, the charging dynamics under PBC can be summarized as follows: 
\begin{itemize}%
    \item Even $N$ (periodicity $E_N(m+N)=E_N(m)$):%
    \begin{itemize}%
        \item At $m=(q+\frac{1}{2})N$ cells are maximally charged, $E_N(m)/N=g$, most excited energy of $H_\text{b}$.%
        \item At $m=qN$, cells are maximally discharged, $E_N(m)/N=0$, GS energy of $H_\text{b}$.%
        \item Otherwise, $E_N(m)/N=g/2$, thus the battery is half charged.%
    \end{itemize}%
    \item Odd $N$ (periodicity $E_N(m+N)=E_N(m)$):%
    \begin{itemize}%
        \item At $m=qN$, cells are maximally discharged, $E_N(m)/N=0$, GS energy of $H_\text{b}$.%
        \item Otherwise, $E_N(m)/N=g/2$, thus the battery is half charged.%
    \end{itemize}%
\end{itemize}%
Additionally, regardless of the parity of $N$, results under open boundary conditions (OBC) behave as their relatives under PBC with odd $N$.

Besides fully characterizing the charging dynamics of the KIC QB under a uniform kicking schedule, it is possible to relax this constraint and instead apply several kicks within a fixed charging time window. This consideration offers greater experimental feasibility, potential enabling an easier implementation on laboratories. Setting $\Delta t_j = t_j - t_{j-1}$ as the time interval between two subsequent kicks $j-1$ and $j$, now we consider the time-evolved unitary operator after $m$ kicks as
\begin{equation}\label{eq:time}
    U(\tau) = e^{-iH_\text{I}(\tau - t_m)}\prod_{j=1}^m e^{-iH_\text{K}\Delta t_j}e^{-iH_\text{I}\Delta t_j},
\end{equation}
with $\tau$ the charging time window and $t_0=0$. Under this setup, we can relate the Floquet operator to the evolution under a static Hamiltonian through the Lie-Trotter product formula. It states that for any two operators $A$ and $B$, we have that $e^{(A+B)\tau} = \lim_{m\to\infty} \left[ e^{A\tau/m}e^{B\tau/m} \right]^m$, which resembles to~\eqref{eq:time} in the limit of applying an infinite number of kicks. Consequently, if we set $\Delta t_j = \Delta t = \tau/m$ as the driving frequency, when $m\to\infty$ (or, equivalently, $\Delta t\to 0$), the KIC converges to its TFIM relative since $\lim_{m\to\infty} U(\tau) = \lim_{m\to\infty} U^m(\tau/m) = e^{-i(H_\text{I} + H_\text{K})\tau}$. This result corresponds to the evolution under the static TFIM Hamiltonian $H_\text{TFIM} = H_\text{I} + H_\text{K}$. It is here where we can actually see the bridge between both models, where in the high-frequency regime periodically-kicked Ising chains converge to the continuously-driven TFIM. This result also holds when non-uniform kicking schedules are considered, which might ease implementation. In fact, in Ref.~\cite{romero2025kicked} a random kicking schedule is tested, observing how for $\tau=1$ and $m \lesssim 10$ kicks, the KIC QB already saturates to the expected energy injected by a continuosly-driven TFIM QB.

More recently in Ref.~\cite{romero2026impact}, the KIC QB setup of Ref.~\cite{romero2025kicked} is studied but including two realistic aspects that might degrade charging performance in laboratories: finite-temperature effects as well as dephasing and thermal dissipation as incoherent contributions. For this purpose, the system is intialized in a Gibbs state of the form
\begin{equation}\label{eq:initial}
    \rho(0) = \frac{e^{-\beta H_\text{th}}}{Z}, \qquad \text{with }Z=\trace{e^{-\beta H_\text{th}}}
\end{equation}
with $\beta=1/T$ the inverse temperature and $H_\text{th} = J_\text{th}\sum_{\braket{ij}} \sigma^x_i\sigma^x_j + h_\text{th}\sum_{i=1}^N \sigma^z_i$ the TFIM Hamiltonian, whose analytical tractability will be leveraged to derive closed-form expressions for the charging dynamics. Notice that while for the infinite-temperature limit we initialize the system under a completely mixed state $\lim_{\beta\to 0} \rho(0) = 1/2^N$, the zero-temperature limit projects to the ground state $\ket{\text{GS}}$ of $H_\text{th}$, i.e., $\lim_{\beta\to\infty} \rho(0) = \proj{\text{GS}}$. This can
also be understood from the formal analogy between the Gibbs operator and
imaginary-time evolution: under the Wick rotation $t\mapsto - i\beta$, the unitary
propagator $e^{-iH_\text{th}t}$ becomes the non-unitary operator
$e^{-\beta H_\text{th}}$. Therefore, as $\beta$ increases, excited-state
contributions are exponentially suppressed and the lowest-energy state
dominates~\cite{simon2005functional}. Remarkably, states in the form of~\eqlabel{eq:initial} are completely passive~\cite{pusz1978passive}, thus no work can be extracted from the Gibbs state at initial times, motivating its inclusion in our studies. Furthermore, we move to a framework where the system evolves under a Lindblad master equation including both dephasing and thermal dissipation as local Lindblad operators in~\eqlabel{eq:lindblad}. 

Dephasing causes the decay of quantum coherences and limits the lifetime of quantum superpositions, as characterized by the dephasing time $T_2$. Amplitude damping induces populations to decay and therefore constraints the lifetime of excitations, as characterized by the relaxation time $T_1$. Both dephasing and amplitude damping might be detrimental for the charging process, as they degrade the quantum resources that can be exploited to enhance charging performance. On the one hand, when thermalization is accounted, it can induce a decay on both populations and coherences present. On the other hand, for a finite-temperature bath, thermal excitations can transfer energy to the system and therefore assist the charging.
The interplay between these two effects and the net effect is governed by temperature, coupling strength, and spectral properties of the bath~\cite{fossfeig2013dynamical}.

Pure dephasing is included through local Lindblad operators $L_i= \sqrt{\gamma_z}\sigma^z_i$, so that~\eqlabel{eq:lindblad} reduces to
\begin{equation}\label{eq:dephasing}
  \frac{\text{d}\rho(t)}{\text{d}t} = -i[H(t), \rho(t)] + \gamma_z\sum_{i=1}^N \left( \sigma^z_i\rho(t)\sigma^z_i - \rho(t) \right).
\end{equation}
Remarkably, when local dephasing is included, the evolution remains quadratic in fermions after applying the Jordan-Wigner transformation to~\eqlabel{eq:dephasing}. Therefore, the system remains integrable even in the presence of dephasing. This feature is leveraged to describe the entire dynamics of the system as a coupled system of $3N/2$ linear differential equations. Their corresponding variables are given by the number operators $n_k=c^\dagger_kc_k$ together with the real and imaginary parts of the anomalous operators $m_k=c_{-k}c_k$ in momentum space. The momentum modes are given by $k\in\{(2n+l)\pi/N, \text{ }\forall n=0,\dots,N-1\}$. Here $l=1$ ($l=0$) correspond to the even-parity (odd-parity) sector. This method follows a two-staged procedure per kick, where the system is first evolved under the coherent part of~\eqlabel{eq:dephasing} and then each kick is applied using conjugations rules.

As a methodological remark, the approach of Ref.~\cite{romero2026impact} can be viewed as an extension of the \emph{third quantization} \cite{prosen2008third}, while allowing the dynamics to be expressed directly in terms of the quadratic variables $\{n_k,\text{Re}[m_k],\text{Im}[m_k]\}$. For local dephasing, the variables form a closed set of $3N/2$ coupled linear differential equations, which provides an efficient description of the full dynamics. Gibbs-state initial conditions can be incorporated by initializing the quadratic variables accordingly. 
\begin{figure}[!tb]
    \centering
    \includegraphics[width=\linewidth]{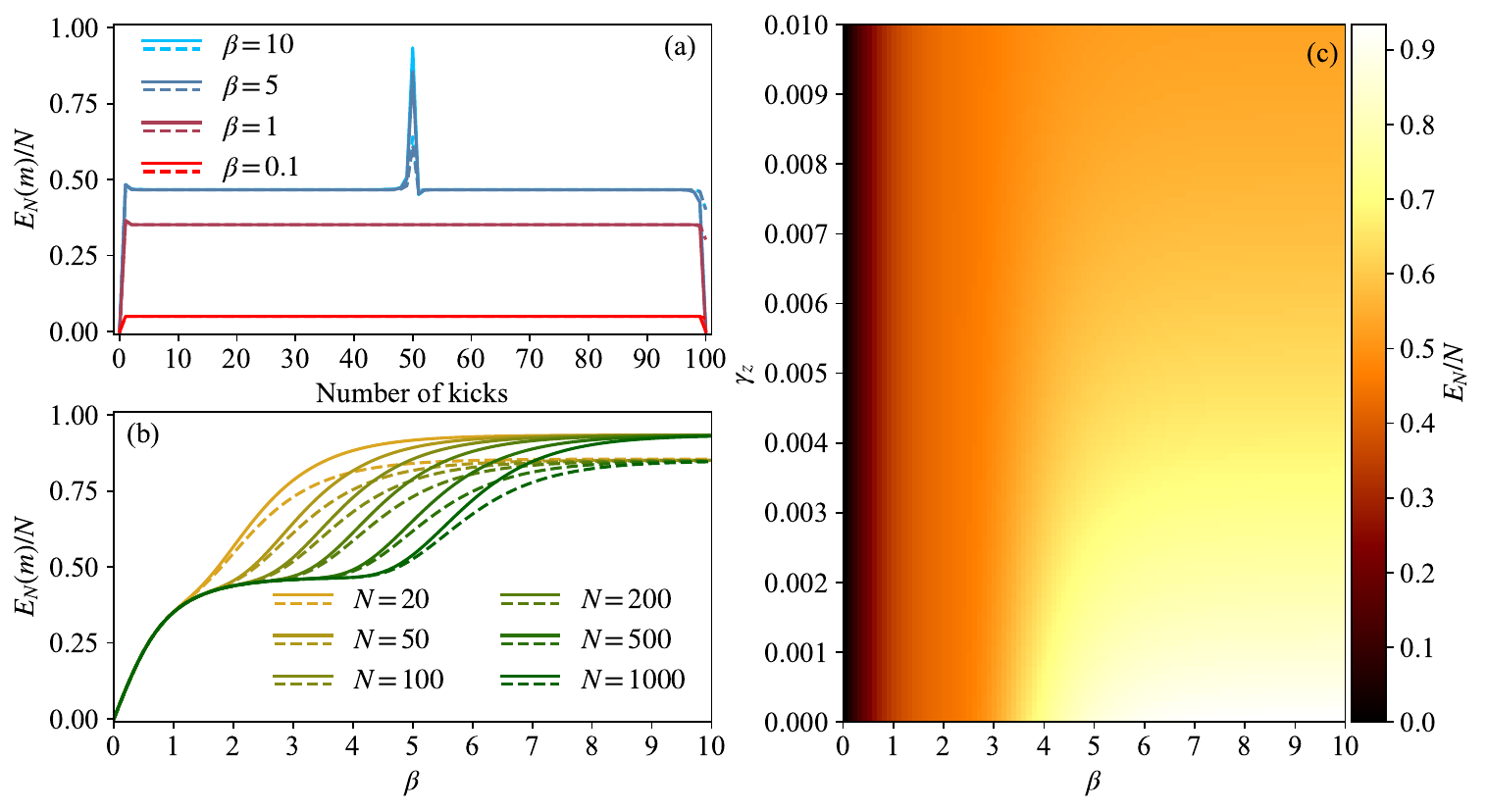}\vspace{-2mm}%
    \caption{Injected energies computed with initial TFIM at $(J_\text{th},h_\text{th}) = (1/2,1)$ and evolved at the self-dual point. (a) Evolution up to $m=100$ kicks for $N=100$ spins with $\gamma_z=0.005$. Results are displayed for different $\beta$ values, ranging from high (red) to low (blue) temperature regimes. (b) Injected energies at $m=N/2$ using $\gamma_z=1/10N$ for various temperatures and system sizes, ranging from smaller (olive) to larger (green) setups. (c) For $N=100$, injected energies at $m=N/2$ kicks for different $(\beta,\gamma_z)$ pairs. Dashed (solid) curves include (discard) dephasing.}\label{fig:dephasing}
\end{figure}%
\begin{figure}[!tb]
    \centering
    \includegraphics[width=\linewidth]{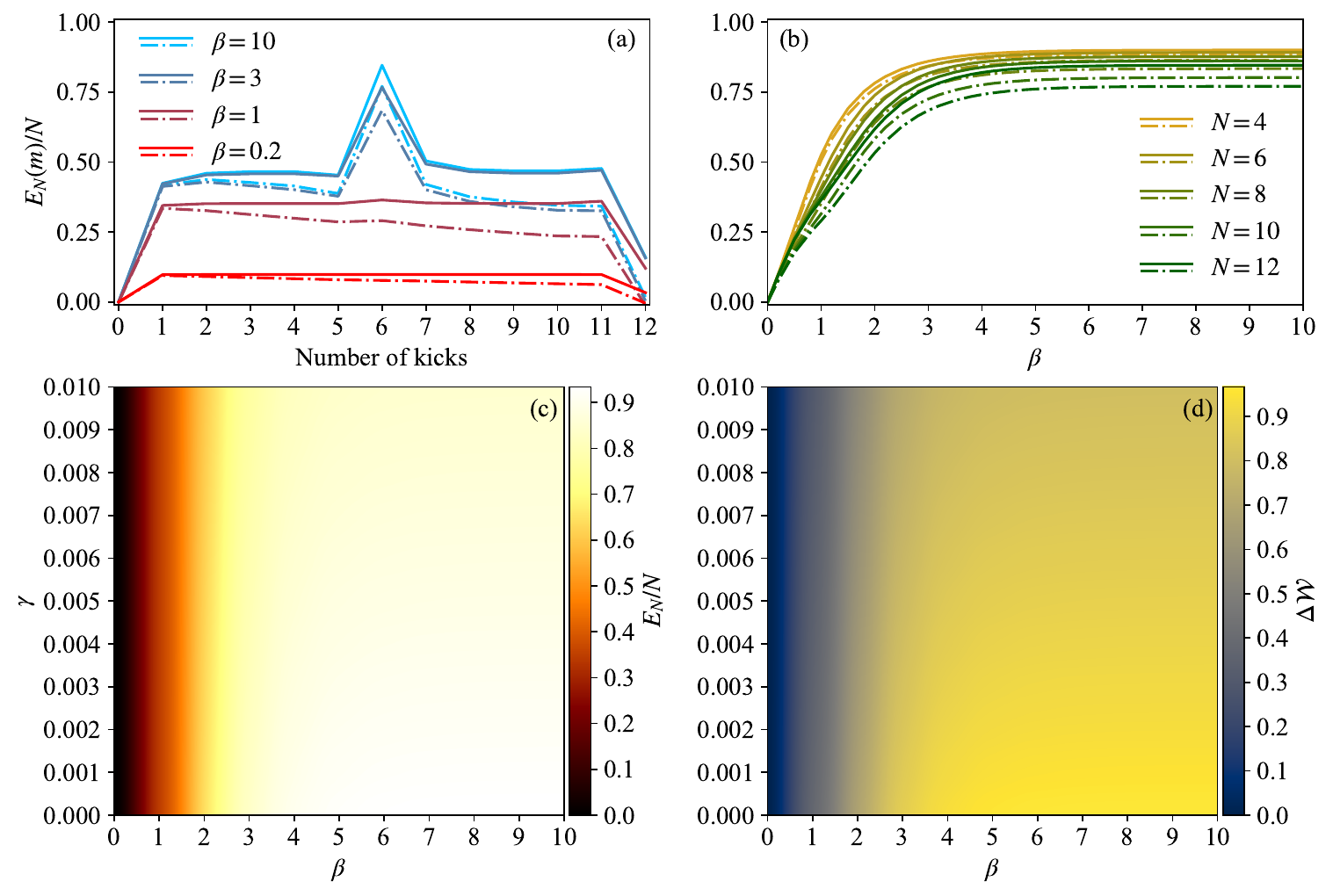}\vspace{-2mm}%
    \caption{(a) Injected energies (solid curves) and ergotropies (dash-dotted) in the presence of dephasing with $\gamma_z=0.01$ up to $m=12$ kicks for a KIC of $N=12$ spins. The initial TFIM is set with $(J_\text{th},h_\text{th}) = (1/2,1)$ and the system evolves under the self-dual point. (b) Same as (a) but for different system sizes $N$ and $\beta$ values, evaluated at $m=N/2$. (c) Injected energies at $m=N/2$ for various $(\beta,\gamma_z)$ pairs, with $N=12$. (d) Same as (c) but for the ergotropy.}\label{fig:erg_dephasing}
\end{figure}%

In Fig.~\ref{fig:dephasing} we compute the normalized injected energies for different temperatures $\beta\in[0,10]$, even system sizes ranging from $N=12$ to $N=1000$, and decoherence rates $\gamma_z\in[0,0.01]$, initializing the state with $(J_\text{th},h_\text{th})=(1/2,1)$. At lower temperatures the protocol can inject energy peaking at $m=N/2$, as expected~\cite{romero2025kicked}, while for $\beta\to0$ the state converges to a completely passive state where no energy is stored. Moreover, with increasing system size and $\gamma_z=0$, the normalized injected energy saturates to the thermodynamic limit value, which is recovered by replacing $\sum_k \mapsto (N/2\pi)\int_{-\pi}^\pi \text{d}k$. In Fig.~\ref{fig:dephasing}c, for a KIC QB of $N=100$ spins at the self-dual point, we show the effect of the initial temperature and decoherence rate on the energy injected at $m=N/2$. As expected, increasing $\beta$ (i.e., approaching zero temperature) increases the injected energy. Regarding decoherence, for large $\gamma_z$ and $\beta$ the normalized injected energy saturates to $1/2$, because the initial state approaches the classical, fully mixed steady state $\rho_\text{ss}=1/2^N$. Coherences decay exponentially as $e^{-\gamma_z m}$, so one can assess robustness by comparing this decay product with the KIC parameters that set the energy scaling: when $\gamma_z m \ll J,b$ the coherent dynamics dominate, whereas for $\gamma_z m \gg J,b$ coherences are strongly suppressed.

To compare how much work can be extracted from the injected energy under decoherence and using~\eqlabel{eq:ergotropy}, we define the normalized injected ergotropy $\Delta\mathcal{W}(\tau)= \mathcal{W}(\rho(\tau)) - \mathcal{W}(\rho(0))$ and numerically compute both quantities up to $N=12$. In~\figlabel{fig:erg_dephasing} we plot them for a KIC QB at the self-dual point, studying the impact of the initial state and decoherence rates. Consistent with the trend observed in~\figlabel{fig:dephasing}, the ergotropy decays with increasing number of kicks.
\begin{figure}[!tb]
    \centering
    \includegraphics[width=\linewidth]{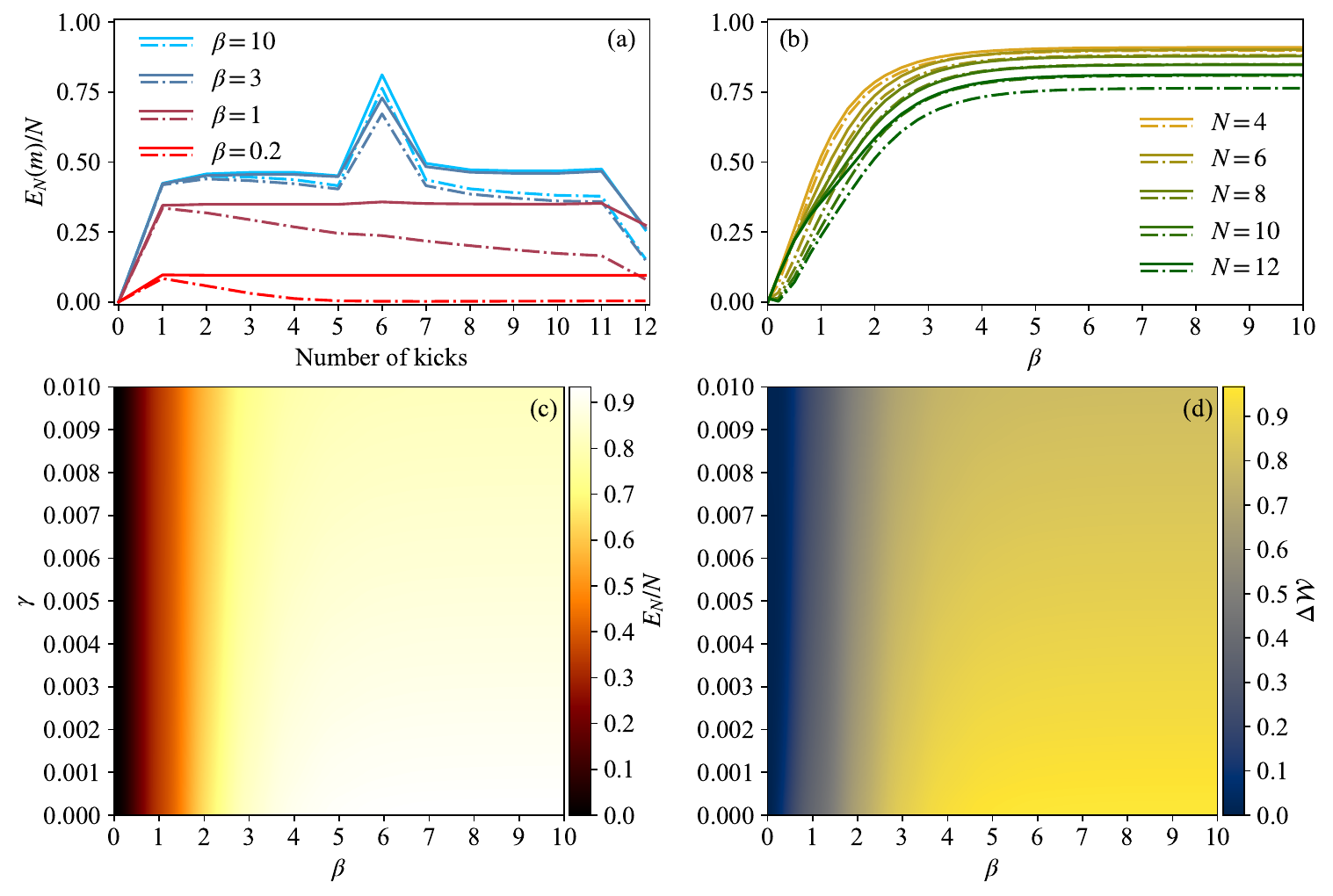}\vspace{-2mm}%
    \caption{(a) Injected energies (solid curves) and ergotropies (dash-dotted) in the presence of thermal dissipation with $\gamma=0.01$ up to $m=12$ kicks for a KIC of $N=12$ spins. The initial TFIM is set with $(J_\text{th},h_\text{th}) = (1/2,1)$ and the system evolves under the self-dual point. (b) Same as (a) but for different system sizes $N$ and $\beta$ values, evaluated at $m=N/2$. (c) Injected energies at $m=N/2$ for various $(\beta,\gamma)$ pairs, with $N=12$. (d) Same as (c) but for the ergotropy.}\label{fig:erg_diss}
\end{figure}%

The thermal dissipation contribution is included as local contributions of the form $L_i = \sqrt{\gamma_-}\sigma^-_i$ and $L_i = \sqrt{\gamma_+}\sigma^+_i$, so that the evolution under~\eqlabel{eq:lindblad} reduces to
\begin{equation}\label{eq:ladder}
\begin{split}
  \frac{\text{d}\rho(t)}{\text{d}t} &= -i[H(t), \rho(t)] + \sum_{s\in\pm}\gamma_s\sum_{i=1}^N \left[ \sigma^s_i\rho(t)\sigma^{s\dagger}_i - \frac{1}{2}\left\{ \sigma^{s\dagger}_i\sigma^s_i, \rho(t)\right\} \right],
\end{split}
\end{equation}
For a finite-temperature bath, thermalization arises from the exchange of energy between the system and the environment, including both emission into the bath and absorption from it. In~\eqlabel{eq:ladder}, this can be captured by $\gamma_- = \gamma ( n_\text{th} + 1 )$ and $\gamma_+ = \gamma n_\text{th}$ with Bose-Einstein occupation number $n_\text{th} = 1/(e^{\beta\omega_0}-1)$, and $\omega_0$ the frequency of the bath mode, setting $\omega_0=1$. The dissipative rates satisfy that $ \gamma_+/\gamma_- = e^{-\beta\omega_0 } $, which is the condition that the ratio of populations meet at equilibrium. Contrary to the previous case, now numerical methods are needed since~\eqlabel{eq:ladder} contains higher-order terms after Jordan-Wigner transformation, disabling its exact diagonalization.%

In~\figlabel{fig:erg_diss} we compute the normalized injected energy and ergotropy for a KIC QB of $N=12$ spins at the self-dual point under thermal dissipation after $m=N/2$ kicks, studying the impact of temperature and dissipation rate $(\beta,\gamma)\in[0,10]\times[0,0.01]$. As expected, lower temperatures and smaller $\gamma$ yield higher injected energies and ergotropies, similar to~\figlabel{fig:erg_dephasing}. Moreover, the charging protocol remains robust against decoherence within the parameter regimes considered, suggesting that it can retain stable performance under the dominant noise mechanisms of current experimental platforms, where dephasing is often a major source of coherence loss in neutral atoms, trapped ions, and solid-state qubits~\cite{romero2026impact}.

\section{Beyond spin-chain architectures: Sachdev-Ye-Kitaev batteries}
\label{sec:syk}

The models discussed so far illustrate the advantages of spin-chain and Floquet-engineered batteries as analytically tractable, controllable, and experimentally accessible platforms for energy storage. A natural question is whether qualitatively different charging mechanisms emerge  beyond spatially ordered and integrable architectures, particularly in strongly interacting
many-body systems with all-to-all random couplings.

The SYK model~\cite{sachdev1993gapless,kitaev2015talk} provides a paradigmatic strongly correlated many-body system consisting of all-to-all connected Majorana fermions with $q$-body random interactions, featuring maximal chaoticity, to address this question. In the large-$N$, low-temperature regime, it exhibits approximate conformal symmetry, fast scrambling, and maximal quantum chaos~\cite{maldacena2016remarks}. These properties have made the SYK model an important analytically tractable framework for exploring connections between strongly correlated quantum matter and holography. Its experimental emulation would therefore provide access to selected many-body signatures associated with holographic systems, rather than constituting a literal realization of a black hole in the laboratory. 

Recently Rossini~\emph{et al.}~\cite{rossini2020quantum} proposed using an SYK Hamiltonian as the charger of a quantum battery. In the direct charging protocol considered there, the battery Hamiltonian is a local sum of spins $H_\text{b} = (g/2)\sum_{i=1}^N \sigma^y_i$, and the charger Hamiltonian is taken as the complex SYK (c-SYK) model
\begin{equation}\label{eq:syk}
    H_\text{c} = \sum_{i,j,k,l=1}^N \mathcal{J}_{ijkl} c^\dagger_i c^\dagger_j c_k c_l.
\end{equation}
The random couplings $\mathcal{J}_{ijkl}$ obey the standard constraints, making $H_\text{c}$ to follow fermionic anticommutation rules and preserve Hermiticity, and they are rescaled to yield a well-defined thermodynamic limit. After applying the Jordan-Wigner transformation, the charger contains highly nonlocal many-body contributions, making it a natural model for collective charging. The c-SYK charger can yield a superextensive charging power scaling $P_N \sim N^{3/2}$. This means that the c-SYK charger plays a role such that collective many-body couplings can enhance charging even in the absence of spatial order.

The physical origin of this scaling can be understood in terms of the correlations generated among different battery cells. Since $H_\text{b}$ is local, the variance of the battery Hamiltonian can be separated into a local and an entangled contribution, whereas the latter can scale superextensively when the charger creates strong many-body correlations that are ultimately responsible of the $N^{3/2}$ scaling of the charging power. 
This identifies collective many-body correlations as the relevant resource behind the charging advantage. Since the SYK model is also a paradigmatic example of a fast scrambler, it is natural to ask whether quantum chaos by itself is also responsible of the charging advantage. Recent analyses bases on standard scrambling diagnostics indicate that it is not the case in a direct sense~\cite{romero2025scrambling}.

Within the broader landscape of QBs, SYK batteries provide a complementary
limit to spin-chain and Floquet-engineered architectures. Spin-chain and kicked-Ising models offer analytically controlled and experimentally motivated
platforms for designing structured charging protocols, whereas SYK batteries
probe strongly collective dynamics generated by disordered all-to-all
interactions. Comparing these regimes helps identify the ingredients that are
essential for quantum charging advantage. In particular, the results suggest
that neither interaction range nor quantum chaos alone is sufficient, and the ability to generate collective correlations among battery cells appears to be central.

\section{Conclusion and outlook}\label{sec:discussion}

QBs have emerged as a promising framework to explore how quantum effects can be exploited for energy storage and transfer at the microscopic scale. This Chapter focuses on spin chain QBs including continuously driven spin chains and Floquet-engineered kicked models, and their connections. Within this framework, we review how many-body interactions, driving protocols and environmental effects influence both energy storage and extractable work.

Spin-chain QBs, initiated by the seminal work of Le \emph{et al.} in 2018, have demonstrated that collective, long-range interactions can unlock a superextensive scaling of charging power, a hallmark of genuine quantum advantage over their classical counterparts. This advantage, however, is sensitive to the interaction range, the presence of disorder, and anisotropy. Understanding and engineering these features has therefore become a central theme, leading to refined criteria for when and how quantum correlations can be harnessed for practical energy storage. 

Rather than aiming primarily at a superextensive charging advantage, the strength of kicked-Ising QBs lies in the analytical control offered by their Floquet structure and exact solvability at particular regimes. Initialization in finite-temperature Gibbs states and the inclusion of dissipative channels further clarify the practical relevance of the kicked-Ising QB. While dephasing and thermal dissipation reduce the stored energy and ergotropy, the protocol displays robustness within the parameter regimes considered. This fact emphasizes that realistic QBs should be assessed not only through injected energy or charging power, but also through extractable work, stability and resilience against environmental couplings.

A central message emerging from this comparison is that continuous and Floquet charging should not be viewed as disconnected strategies. Continuously driven spin chains offer a natural route to many-body charging through static or slowly varying Hamiltonians, while kicked models implement the same physical ingredients in a stroboscopic and gate-like form. The kicked-Ising QB makes this connection particularly transparent: by tuning the kicking schedule, one can interpolate between periodically driven Floquet dynamics and the continuously driven transverse-field Ising limit. In this sense, Floquet engineering provides not only an alternative charging protocol, but also a controlled language to connect analytical solvability, digital implementability, and experimentally relevant driving schemes.

The SYK QBs represent a complementary limit to the spin-chain and kicked-Ising aforementioned architectures. Instead of relying on spatially ordered or Floquet-engineered dynamics, they exploit strongly collective, all-to-all random interactions. Their charging power $P_N\sim N^{3/2}$ scaling highlights the role of many-body correlations in enhancing charging performance. Recent studies suggest, however, that this enhancement is not a consequence of chaos or fast scrambling.

Looking ahead, several directions appear particularly important. First, it remains necessary to clarify what should be meant by a useful quantum advantage in QBs, beyond the scaling of the charging power alone. Quantities such as ergotropy, stability, self-discharging and robustness against environmental couplings should play a central role in assessment. Second, Floquet-engineered QBs provide a promising framework in which driving protocols can be designed, optimized and connected to gate-based device implementations, but their performance must be tested under finite coherence times, cross-talk, calibration errors and limited connectivity. Finally, the impact of open-system effects should not be regarded as detrimental by default. Dephasing, relaxation and thermalization can also be used to identify regimes of robustness or to assist energy transfer through bath engineering. Combining analytical tractability, numerical optimization and experimental constraints will therefore be essential for moving from idealized charging protocols toward practical and controllable quantum energy-storage devices.

\begin{acknowledgement}
This work is supported by the Spanish Ministry of Science, Innovation, and Universities under Grants PID2024-157842OA-I00 and PID2021-126273NB-I00 funded by MCIN/AEI/10.13039/501100011033 and by ``ERDF A way of making Europe'' and ``ERDF Invest in your Future'', Spanish national project in the field of Artificial Intelligence (AIA2025-163435-C44), and the Severo Ochoa Centres of Excellence program through Grant CEX2024-001445-S.
\end{acknowledgement}

\ethics{Competing Interests}{The authors have no conflicts of interest to declare that are relevant to the content of this chapter.}

\bibliographystyle{spphys}
\bibliography{biblio}

\begin{thebibliography}{10}
\providecommand{\url}[1]{{#1}}
\providecommand{\urlprefix}{URL }
\expandafter\ifx\csname urlstyle\endcsname\relax
  \providecommand{\doi}[1]{DOI \discretionary{}{}{}#1}\else
  \providecommand{\doi}{DOI \discretionary{}{}{}\begingroup \urlstyle{rm}\Url}\fi

\bibitem{alicki2013entanglement}
R.~Alicki, M.~Fannes, Phys. Rev. E \textbf{87}, 042123 (2013).
\newblock \doi{10.1103/PhysRevE.87.042123}.
\newblock \urlprefix\url{https://link.aps.org/doi/10.1103/PhysRevE.87.042123}

\bibitem{hovhannisyan2013entanglement}
K.V. Hovhannisyan, M.~Perarnau-Llobet, M.~Huber, A.~Ac\'{\i}n, Phys. Rev. Lett. \textbf{111}, 240401 (2013).
\newblock \doi{10.1103/PhysRevLett.111.240401}.
\newblock \urlprefix\url{https://link.aps.org/doi/10.1103/PhysRevLett.111.240401}

\bibitem{binder2015quantacell}
F.C. Binder, S.~Vinjanampathy, K.~Modi, J.~Goold, New J. Phys. \textbf{17}(7), 075015 (2015).
\newblock \doi{10.1088/1367-2630/17/7/075015}.
\newblock \urlprefix\url{https://doi.org/10.1088/1367-2630/17/7/075015}

\bibitem{hu2022optimal}
C.K. Hu, J.~Qiu, P.J.P. Souza, J.~Yuan, Y.~Zhou, L.~Zhang, J.~Chu, X.~Pan, L.~Hu, J.~Li, Y.~Xu, Y.~Zhong, S.~Liu, F.~Yan, D.~Tan, R.~Bachelard, C.J. Villas-Boas, A.C. Santos, D.~Yu, Quantum Science and Technology \textbf{7}(4), 045018 (2022).
\newblock \doi{10.1088/2058-9565/ac8444}.
\newblock \urlprefix\url{https://doi.org/10.1088/2058-9565/ac8444}

\bibitem{hu2026quantumcharging}
C.K. Hu, C.~Liu, J.~Zhao, L.~Zhong, Y.~Zhou, M.~Liu, H.~Yuan, Y.~Lin, Y.~Xu, G.~Hu, G.~Xie, Z.~Liu, R.~Zhou, Y.~Ri, W.~Zhang, R.~Deng, A.~Saguia, X.~Linpeng, M.S. Sarandy, S.~Liu, A.C. Santos, D.~Tan, D.~Yu, Phys. Rev. Lett. \textbf{136}, 060401 (2026).
\newblock \doi{10.1103/sp5l-c6m8}.
\newblock \urlprefix\url{https://link.aps.org/doi/10.1103/sp5l-c6m8}

\bibitem{wenniger2023experimental}
I.~Maillette~de Buy~Wenniger, S.E. Thomas, M.~Maffei, S.C. Wein, M.~Pont, N.~Belabas, S.~Prasad, A.~Harouri, A.~Lema\^{\i}tre, I.~Sagnes, N.~Somaschi, A.~Auff\`eves, P.~Senellart, Phys. Rev. Lett. \textbf{131}, 260401 (2023).
\newblock \doi{10.1103/PhysRevLett.131.260401}.
\newblock \urlprefix\url{https://link.aps.org/doi/10.1103/PhysRevLett.131.260401}

\bibitem{quach2022superabsorption}
J.Q. Quach, K.E. McGhee, L.~Ganzer, D.M. Rouse, B.W. Lovett, E.M. Gauger, J.~Keeling, G.~Cerullo, D.G. Lidzey, T.~Virgili, Science Advances \textbf{8}(2), eabk3160 (2022).
\newblock \doi{10.1126/sciadv.abk3160}.
\newblock \urlprefix\url{https://www.science.org/doi/abs/10.1126/sciadv.abk3160}

\bibitem{joshi2022experimental}
J.~Joshi, T.S. Mahesh, Phys. Rev. A \textbf{106}, 042601 (2022).
\newblock \doi{10.1103/PhysRevA.106.042601}.
\newblock \urlprefix\url{https://link.aps.org/doi/10.1103/PhysRevA.106.042601}

\bibitem{campaioli2017enhancing}
F.~Campaioli, F.A. Pollock, F.C. Binder, L.~C\'eleri, J.~Goold, S.~Vinjanampathy, K.~Modi, Phys. Rev. Lett. \textbf{118}, 150601 (2017).
\newblock \doi{10.1103/PhysRevLett.118.150601}.
\newblock \urlprefix\url{https://link.aps.org/doi/10.1103/PhysRevLett.118.150601}

\bibitem{farre2020bounds}
S.~Juli\`a-Farr\'e, T.~Salamon, A.~Riera, M.N. Bera, M.~Lewenstein, Phys. Rev. Res. \textbf{2}, 023113 (2020).
\newblock \doi{10.1103/PhysRevResearch.2.023113}.
\newblock \urlprefix\url{https://link.aps.org/doi/10.1103/PhysRevResearch.2.023113}

\bibitem{gyhm2022quantum}
J.Y. Gyhm, D.~\v{S}afr\'anek, D.~Rosa, Phys. Rev. Lett. \textbf{128}, 140501 (2022).
\newblock \doi{10.1103/PhysRevLett.128.140501}.
\newblock \urlprefix\url{https://link.aps.org/doi/10.1103/PhysRevLett.128.140501}

\bibitem{ferraro2018highpower}
D.~Ferraro, M.~Campisi, G.M. Andolina, V.~Pellegrini, M.~Polini, Phys. Rev. Lett. \textbf{120}, 117702 (2018).
\newblock \doi{10.1103/PhysRevLett.120.117702}.
\newblock \urlprefix\url{https://link.aps.org/doi/10.1103/PhysRevLett.120.117702}

\bibitem{andolina2025genuine}
G.M. Andolina, V.~Stanzione, V.~Giovannetti, M.~Polini, Phys. Rev. Lett. \textbf{134}, 240403 (2025).
\newblock \doi{10.1103/kzvn-dj7v}.
\newblock \urlprefix\url{https://link.aps.org/doi/10.1103/kzvn-dj7v}

\bibitem{le2018spinchain}
T.P. Le, J.~Levinsen, K.~Modi, M.M. Parish, F.A. Pollock, Phys. Rev. A \textbf{97}, 022106 (2018).
\newblock \doi{10.1103/PhysRevA.97.022106}.
\newblock \urlprefix\url{https://link.aps.org/doi/10.1103/PhysRevA.97.022106}

\bibitem{rossini2019manybody}
D.~Rossini, G.M. Andolina, M.~Polini, Phys. Rev. B \textbf{100}, 115142 (2019).
\newblock \doi{10.1103/PhysRevB.100.115142}.
\newblock \urlprefix\url{https://link.aps.org/doi/10.1103/PhysRevB.100.115142}

\bibitem{ghosh2020enhancement}
S.~Ghosh, T.~Chanda, A.~Sen(De), Phys. Rev. A \textbf{101}, 032115 (2020).
\newblock \doi{10.1103/PhysRevA.101.032115}.
\newblock \urlprefix\url{https://link.aps.org/doi/10.1103/PhysRevA.101.032115}

\bibitem{arjmandi2022enhancing}
M.B. Arjmandi, H.~Mohammadi, A.C. Santos, Phys. Rev. E \textbf{105}, 054115 (2022).
\newblock \doi{10.1103/PhysRevE.105.054115}.
\newblock \urlprefix\url{https://link.aps.org/doi/10.1103/PhysRevE.105.054115}

\bibitem{ghosh2022dimensional}
S.~Ghosh, A.~Sen(De), Phys. Rev. A \textbf{105}, 022628 (2022).
\newblock \doi{10.1103/PhysRevA.105.022628}.
\newblock \urlprefix\url{https://link.aps.org/doi/10.1103/PhysRevA.105.022628}

\bibitem{ali2024ergotropy}
A.~Ali, S.~Al-Kuwari, M.I. Hussain, T.~Byrnes, M.T. Rahim, J.Q. Quach, M.~Ghominejad, S.~Haddadi, Phys. Rev. A \textbf{110}, 052404 (2024).
\newblock \doi{10.1103/PhysRevA.110.052404}.
\newblock \urlprefix\url{https://link.aps.org/doi/10.1103/PhysRevA.110.052404}

\bibitem{bhattacharya2026heisenberg}
S.~Bhattacharya, V.B. Sabale, A.~Kumar, New Journal of Physics \textbf{28}(1), 014508 (2026).
\newblock \doi{10.1088/1367-2630/ae33e4}.
\newblock \urlprefix\url{https://doi.org/10.1088/1367-2630/ae33e4}

\bibitem{huangfu2021highcapacity}
Y.~Huangfu, J.~Jing, Phys. Rev. E \textbf{104}, 024129 (2021).
\newblock \doi{10.1103/PhysRevE.104.024129}.
\newblock \urlprefix\url{https://link.aps.org/doi/10.1103/PhysRevE.104.024129}

\bibitem{gao2022scaling}
L.~Gao, C.~Cheng, W.B. He, R.~Mondaini, X.W. Guan, H.Q. Lin, Phys. Rev. Res. \textbf{4}, 043150 (2022).
\newblock \doi{10.1103/PhysRevResearch.4.043150}.
\newblock \urlprefix\url{https://link.aps.org/doi/10.1103/PhysRevResearch.4.043150}

\bibitem{grazi2024controlling}
R.~Grazi, D.~Sacco~Shaikh, M.~Sassetti, N.~Traverso~Ziani, D.~Ferraro, Phys. Rev. Lett. \textbf{133}, 197001 (2024).
\newblock \doi{10.1103/PhysRevLett.133.197001}.
\newblock \urlprefix\url{https://link.aps.org/doi/10.1103/PhysRevLett.133.197001}

\bibitem{catalano2024frustrating}
A.~Catalano, S.~Giampaolo, O.~Morsch, V.~Giovannetti, F.~Franchini, PRX Quantum \textbf{5}, 030319 (2024).
\newblock \doi{10.1103/PRXQuantum.5.030319}.
\newblock \urlprefix\url{https://link.aps.org/doi/10.1103/PRXQuantum.5.030319}

\bibitem{dou2022charging}
F.Q. Dou, Y.J. Wang, J.A. Sun.
\newblock {Charging advantages of Lipkin-Meshkov-Glick quantum battery} (2022).
\newblock Preprint at \url{https://arxiv.org/abs/2208.04831}

\bibitem{schmid2026superextensive}
H.~Schmid, F.~von Oppen, G.~Refael, Y.~Peng, Phys. Rev. B \textbf{114}, 014313 (2026).
\newblock \doi{10.1103/pkc6-c3pd}.
\newblock \urlprefix\url{https://link.aps.org/doi/10.1103/pkc6-c3pd}

\bibitem{ho2026boosting}
L.B. Ho, D.T. Hoang, T.D. Anh-Tai, T.~Busch, T.~Fogarty.
\newblock {Boosting the Performance of a Lipkin-Meshkov-Glick Quantum Battery via Symmetry-Breaking Quenches and Bosonic Baths} (2026).
\newblock Preprint at \url{https://arxiv.org/abs/2602.17121}

\bibitem{grazi2025charging}
R.~Grazi, F.~Cavaliere, M.~Sassetti, D.~Ferraro, N.~{Traverso Ziani}, Chaos, Solitons \& Fractals \textbf{196}, 116383 (2025).
\newblock \doi{10.1016/j.chaos.2025.116383}.
\newblock \urlprefix\url{https://www.sciencedirect.com/science/article/pii/S0960077925003960}

\bibitem{pavone2026cluster}
A.~Pavone, F.L. Cavagnaro, M.~Carrega, R.~Grazi, D.~Ferraro, N.~Traverso~Ziani, Phys. Rev. B \textbf{114}, 065405 (2026).
\newblock \doi{10.1103/rdhw-9kh3}.
\newblock \urlprefix\url{https://link.aps.org/doi/10.1103/rdhw-9kh3}

\bibitem{zhao2025nonmarkovian}
S.C. Zhao, Z.R. Zhao, N.Y. Zhuang, Phys. Rev. E \textbf{112}, 024129 (2025).
\newblock \doi{10.1103/xqtv-qbyk}.
\newblock \urlprefix\url{https://link.aps.org/doi/10.1103/xqtv-qbyk}

\bibitem{dou2022cavity}
F.Q. Dou, H.~Zhou, J.A. Sun, Phys. Rev. A \textbf{106}, 032212 (2022).
\newblock \doi{10.1103/PhysRevA.106.032212}.
\newblock \urlprefix\url{https://link.aps.org/doi/10.1103/PhysRevA.106.032212}

\bibitem{sun2025cavity}
P.Y. Sun, H.~Zhou, F.Q. Dou, New Journal of Physics \textbf{27}(12), 124513 (2025).
\newblock \doi{10.1088/1367-2630/ae2a62}.
\newblock \urlprefix\url{https://doi.org/10.1088/1367-2630/ae2a62}

\bibitem{zhao2021quantum}
F.~Zhao, F.Q. Dou, Q.~Zhao, Phys. Rev. A \textbf{103}, 033715 (2021).
\newblock \doi{10.1103/PhysRevA.103.033715}.
\newblock \urlprefix\url{https://link.aps.org/doi/10.1103/PhysRevA.103.033715}

\bibitem{chang2021optimal}
W.~Chang, T.R. Yang, H.~Dong, L.~Fu, X.~Wang, Y.Y. Zhang, New Journal of Physics \textbf{23}(10), 103026 (2021).
\newblock \doi{10.1088/1367-2630/ac2a5b}.
\newblock \urlprefix\url{https://doi.org/10.1088/1367-2630/ac2a5b}

\bibitem{yao2022optimal}
Y.~Yao, X.Q. Shao, Phys. Rev. E \textbf{106}, 014138 (2022).
\newblock \doi{10.1103/PhysRevE.106.014138}.
\newblock \urlprefix\url{https://link.aps.org/doi/10.1103/PhysRevE.106.014138}

\bibitem{fajar2025noise}
M.M. Fajar, B.N. Pratiwi, S.~Pramono, G.K. Sunnardianto, A.R.T. Nugraha.
\newblock {Noise-induced Zeno-like effect in a spin-chain quantum battery} (2025).
\newblock Preprint at \url{https://arxiv.org/abs/2509.01603}

\bibitem{liu2021entanglement}
J.X. Liu, H.L. Shi, Y.H. Shi, X.H. Wang, W.L. Yang, Phys. Rev. B \textbf{104}, 245418 (2021).
\newblock \doi{10.1103/PhysRevB.104.245418}.
\newblock \urlprefix\url{https://link.aps.org/doi/10.1103/PhysRevB.104.245418}

\bibitem{peng2021lower}
L.~Peng, W.B. He, S.~Chesi, H.Q. Lin, X.W. Guan, Phys. Rev. A \textbf{103}, 052220 (2021).
\newblock \doi{10.1103/PhysRevA.103.052220}.
\newblock \urlprefix\url{https://link.aps.org/doi/10.1103/PhysRevA.103.052220}

\bibitem{qi2021magnon}
S.f. Qi, J.~Jing, Phys. Rev. A \textbf{104}, 032606 (2021).
\newblock \doi{10.1103/PhysRevA.104.032606}.
\newblock \urlprefix\url{https://link.aps.org/doi/10.1103/PhysRevA.104.032606}

\bibitem{barra2022quantum}
F.~Barra, K.V. Hovhannisyan, A.~Imparato, New Journal of Physics \textbf{24}(1), 015003 (2022).
\newblock \doi{10.1088/1367-2630/ac43ed}.
\newblock \urlprefix\url{https://doi.org/10.1088/1367-2630/ac43ed}

\bibitem{guo2024analytically}
W.X. Guo, F.M. Yang, F.Q. Dou, Phys. Rev. A \textbf{109}, 032201 (2024).
\newblock \doi{10.1103/PhysRevA.109.032201}.
\newblock \urlprefix\url{https://link.aps.org/doi/10.1103/PhysRevA.109.032201}

\bibitem{romero2025kicked}
S.V. Romero, X.~Chen, Y.~Ban, Phys. Rev. Lett. \textbf{137}, 050404 (2026).
\newblock \doi{10.1103/j35s-xv5k}.
\newblock \urlprefix\url{https://link.aps.org/doi/10.1103/j35s-xv5k}

\bibitem{mondal2022periodically}
S.~Mondal, S.~Bhattacharjee, Phys. Rev. E \textbf{105}, 044125 (2022).
\newblock \doi{10.1103/PhysRevE.105.044125}.
\newblock \urlprefix\url{https://link.aps.org/doi/10.1103/PhysRevE.105.044125}

\bibitem{pg2025dichotomy}
S.~P.~G., J.B. Kannan, S.H. Tekur, M.S. Santhanam, Phys. Rev. B \textbf{111}, 054314 (2025).
\newblock \doi{10.1103/PhysRevB.111.054314}.
\newblock \urlprefix\url{https://link.aps.org/doi/10.1103/PhysRevB.111.054314}

\bibitem{puri2025floquet}
S.~Puri, T.K. Konar, L.G.C. Lakkaraju, A.S. De.
\newblock Floquet driven long-range interactions induce super-extensive scaling in quantum batteries (2025).
\newblock Preprint at \url{https://arxiv.org/abs/2412.00921}

\bibitem{romero2026impact}
S.V. Romero, X.~Chen, Y.~Ban.
\newblock Impact of thermal and dissipative effects in a periodically-kicked quantum battery (2026).
\newblock Preprint at \url{https://arxiv.org/abs/2604.24409}

\bibitem{zaletel2023quantum}
M.P. Zaletel, M.~Lukin, C.~Monroe, C.~Nayak, F.~Wilczek, N.Y. Yao, Rev. Mod. Phys. \textbf{95}, 031001 (2023).
\newblock \doi{10.1103/RevModPhys.95.031001}.
\newblock \urlprefix\url{https://link.aps.org/doi/10.1103/RevModPhys.95.031001}

\bibitem{sachdev1993gapless}
S.~Sachdev, J.~Ye, Phys. Rev. Lett. \textbf{70}, 3339 (1993).
\newblock \doi{10.1103/PhysRevLett.70.3339}.
\newblock \urlprefix\url{https://link.aps.org/doi/10.1103/PhysRevLett.70.3339}

\bibitem{kitaev2015talk}
A.~Kitaev.
\newblock {\emph{A simple model of quantum holography.} Talks at KITP} (2015).
\newblock {URL}~\url{http://online.kitp.ucsb.edu/online/entangled15/kitaev/}, \url{http://online.kitp.ucsb.edu/online/entangled15/kitaev2/}

\bibitem{maldacena2016remarks}
J.~Maldacena, D.~Stanford, Phys. Rev. D \textbf{94}, 106002 (2016).
\newblock \doi{10.1103/PhysRevD.94.106002}.
\newblock \urlprefix\url{https://link.aps.org/doi/10.1103/PhysRevD.94.106002}

\bibitem{rosa2020ultrastable}
D.~Rosa, D.~Rossini, G.M. Andolina, M.~Polini, M.~Carrega, Journal of High Energy Physics \textbf{2020}(11), 67 (2020).
\newblock \doi{10.1007/JHEP11(2020)067}.
\newblock \urlprefix\url{https://doi.org/10.1007/JHEP11(2020)067}

\bibitem{rossini2020quantum}
D.~Rossini, G.M. Andolina, D.~Rosa, M.~Carrega, M.~Polini, Phys. Rev. Lett. \textbf{125}, 236402 (2020).
\newblock \doi{10.1103/PhysRevLett.125.236402}.
\newblock \urlprefix\url{https://link.aps.org/doi/10.1103/PhysRevLett.125.236402}

\bibitem{kim2022operator}
J.~Kim, J.~Murugan, J.~Olle, D.~Rosa, Phys. Rev. A \textbf{105}, L010201 (2022).
\newblock \doi{10.1103/PhysRevA.105.L010201}.
\newblock \urlprefix\url{https://link.aps.org/doi/10.1103/PhysRevA.105.L010201}

\bibitem{romero2025scrambling}
S.V. Romero, Y.~Ding, X.~Chen, Y.~Ban, Journal of High Energy Physics \textbf{2025}(5), 21 (2025).
\newblock \doi{10.1007/JHEP05(2025)021}.
\newblock \urlprefix\url{https://doi.org/10.1007/JHEP05(2025)021}

\bibitem{divi2025syk}
F.~Divi, J.~Murugan, D.~Rosa, Phys. Rev. B \textbf{111}, 075138 (2025).
\newblock \doi{10.1103/PhysRevB.111.075138}.
\newblock \urlprefix\url{https://link.aps.org/doi/10.1103/PhysRevB.111.075138}

\bibitem{campaioli2024colloquium}
F.~Campaioli, S.~Gherardini, J.Q. Quach, M.~Polini, G.M. Andolina, Rev. Mod. Phys. \textbf{96}, 031001 (2024).
\newblock \doi{10.1103/RevModPhys.96.031001}.
\newblock \urlprefix\url{https://link.aps.org/doi/10.1103/RevModPhys.96.031001}

\bibitem{allahverdyan2004maximal}
A.E. Allahverdyan, R.~Balian, T.M. Nieuwenhuizen, Europhysics Letters \textbf{67}(4), 565 (2004).
\newblock \doi{10.1209/epl/i2004-10101-2}.
\newblock \urlprefix\url{https://dx.doi.org/10.1209/epl/i2004-10101-2}

\bibitem{pusz1978passive}
W.~Pusz, S.L. Woronowicz, Commun. Math. Phys. \textbf{58}(3), 273–290 (1978).
\newblock \doi{10.1007/BF01614224}.
\newblock \urlprefix\url{http://link.springer.com/10.1007/BF01614224}

\bibitem{santos2019stable}
A.C. Santos, B.~\c{C}akmak, S.~Campbell, N.T. Zinner, Phys. Rev. E \textbf{100}, 032107 (2019).
\newblock \doi{10.1103/PhysRevE.100.032107}.
\newblock \urlprefix\url{https://link.aps.org/doi/10.1103/PhysRevE.100.032107}

\bibitem{lieb1961two}
E.~Lieb, T.~Schultz, D.~Mattis, Ann. Phys. \textbf{16}(3), 407 (1961).
\newblock \doi{https://doi.org/10.1016/0003-4916(61)90115-4}.
\newblock \urlprefix\url{https://www.sciencedirect.com/science/article/pii/0003491661901154}

\bibitem{rozenbaum2017lyapunov}
E.B. Rozenbaum, S.~Ganeshan, V.~Galitski, Phys. Rev. Lett. \textbf{118}, 086801 (2017).
\newblock \doi{10.1103/PhysRevLett.118.086801}.
\newblock \urlprefix\url{https://link.aps.org/doi/10.1103/PhysRevLett.118.086801}

\bibitem{waltner2021localization}
D.~Waltner, P.~Braun, Phys. Rev. B \textbf{104}, 054432 (2021).
\newblock \doi{10.1103/PhysRevB.104.054432}.
\newblock \urlprefix\url{https://link.aps.org/doi/10.1103/PhysRevB.104.054432}

\bibitem{Akila_2016}
M.~Akila, D.~Waltner, B.~Gutkin, T.~Guhr, Journal of Physics A: Mathematical and Theoretical \textbf{49}(37), 375101 (2016).
\newblock \doi{10.1088/1751-8113/49/37/375101}.
\newblock \urlprefix\url{https://doi.org/10.1088/1751-8113/49/37/375101}

\bibitem{kim2023evidence}
Y.~Kim, A.~Eddins, S.~Anand, K.X. Wei, E.~Van Den~Berg, S.~Rosenblatt, H.~Nayfeh, Y.~Wu, M.~Zaletel, K.~Temme, A.~Kandala, Nature \textbf{618}(7965), 500 (2023).
\newblock \doi{10.1038/s41586-023-06096-3}.
\newblock \urlprefix\url{https://www.nature.com/articles/s41586-023-06096-3}

\bibitem{miessen2024benchmarking}
A.~Miessen, D.J. Egger, I.~Tavernelli, G.~Mazzola, PRX Quantum \textbf{5}, 040320 (2024).
\newblock \doi{10.1103/PRXQuantum.5.040320}.
\newblock \urlprefix\url{https://link.aps.org/doi/10.1103/PRXQuantum.5.040320}

\bibitem{visuri2026digitized}
A.M. Visuri, A.~Gomez~Cadavid, B.A. Bhargava, S.V. Romero, A.~Grabarits, P.~Chandarana, E.~Solano, A.~Del~Campo, N.N. Hegade, npj Quantum Information \textbf{12}(1), 47 (2026).
\newblock \doi{10.1038/s41534-026-01208-z}.
\newblock \urlprefix\url{https://www.nature.com/articles/s41534-026-01208-z}

\bibitem{fischer2026dynamical}
L.E. Fischer, M.~Leahy, A.~Eddins, N.~Keenan, D.~Ferracin, M.A.C. Rossi, Y.~Kim, A.~He, F.~Pietracaprina, B.~Sokolov, S.~Dooley, Z.~Zimborás, F.~Tacchino, S.~Maniscalco, J.~Goold, G.~García-Pérez, I.~Tavernelli, A.~Kandala, S.N. Filippov, Nature Physics \textbf{22}(2), 302 (2026).
\newblock \doi{10.1038/s41567-025-03144-9}.
\newblock \urlprefix\url{https://www.nature.com/articles/s41567-025-03144-9}

\bibitem{shinjo2026unveiling}
K.~Shinjo, K.~Seki, T.~Shirakawa, R.Y. Sun, S.~Yunoki, npj Quantum Information \textbf{12}(1), 41 (2026).
\newblock \doi{10.1038/s41534-026-01193-3}.
\newblock \urlprefix\url{https://www.nature.com/articles/s41534-026-01193-3}

\bibitem{simon2005functional}
B.~Simon, in \emph{{Functional Integration and Quantum Physics}} (AMS Chelsea Publishing, Providence, RI, 2005), chap.~II, pp. 45--70.
\newblock \urlprefix\url{https://bookstore.ams.org/view?ProductCode=CHEL/351.H}

\bibitem{fossfeig2013dynamical}
M.~Foss-Feig, K.R.A. Hazzard, J.J. Bollinger, A.M. Rey, C.W. Clark, New J. Phys. \textbf{15}(11), 113008 (2013).
\newblock \doi{10.1088/1367-2630/15/11/113008}.
\newblock \urlprefix\url{https://doi.org/10.1088/1367-2630/15/11/113008}

\bibitem{prosen2008third}
T.~Prosen, New Journal of Physics \textbf{10}(4), 043026 (2008).
\newblock \doi{10.1088/1367-2630/10/4/043026}.
\newblock \urlprefix\url{https://doi.org/10.1088/1367-2630/10/4/043026}

\end{thebibliography}

\end{document}